# Ideal band structures for high-performance thermoelectric materials with band convergence

Yuya Hattori[1,*], Hidetomo Usui[2] and Yoshikazu Mizuguchi[1]

[1] *Department of Physics, Tokyo Metropolitan University, Tokyo 192-0397, Japan*

[2] *Department of Engineering Science, The University of Electro-Communications, Tokyo 182-8585, Japan*

[*] Correspondence and requests for materials should be addressed to Yuya Hattori (yhattori@tmu.ac.jp)

## Abstract

We investigate optimal band structures in band-converged systems to achieve high thermoelectric figure of merit, $zT$, using numerical calculations based on Boltzmann transport theory. Using a two parabolic band model with independently tunable band parameters, we derive quantitative and practically useful design principles for thermoelectric materials. The main conclusions are as follows: (i) To suppress the bipolar effect, a band gap $E_g$ satisfying $E_g \geq 5k_BT$ is required for the Seebeck coefficient $S$, and $E_g \geq 7k_BT$ is required for the electronic thermal conductivity $\kappa_{el}$. The $5k_BT$ criterion for $S$ is supported by experimental data for representative thermoelectric materials, which broadly follow $T_{BP} = E_g/(5k_B)$, where $T_{BP}$ denotes the onset temperature of bipolar transport; (ii) In band-converged systems, the energy separation between the band edges $\Delta E$ should satisfy $\Delta E \approx 0$ to maximize $zT$ when interband scattering is insignificant; (iii) The optimal chemical potential $\mu$ is governed by the balance between electronic thermal conductivity $\kappa_{el}$ and lattice thermal conductivity $\kappa_{lat}$, shifting toward the band gap as $\kappa_{lat}$ decreases; (iv) Achieving high spectral conductivity $\Sigma$ (high band degeneracy $N$, density of states effective mass $m^*_{DOS}$, and relaxation time $\tau$) near the band edge is essential for achieving high $zT$.

## 1. Introduction

Thermoelectric materials enable the direct conversion of waste heat into electricity. Thermoelectric energy conversion devices are particularly attractive for applications requiring long operational lifetimes with minimal maintenance, such as those in space environments. Achieving high thermoelectric conversion efficiency requires large values of the figure of merit $zT$, defined as $zT = \sigma S^2 T/(\kappa_{el} + \kappa_{lat})$, where $\sigma$, $S$, $T$, $\kappa_{el}$, and $\kappa_{lat}$ represent the electrical conductivity, Seebeck coefficient, absolute temperature, electronic thermal conductivity, and lattice thermal conductivity, respectively [1]. Despite extensive experimental and theoretical efforts, improvement of $zT$ remains challenging because these transport parameters, especially $\sigma$ and $S$, are strongly interdependent.

From a band-structure perspective, the transport distribution function $\Sigma(E,T)$, or the spectral conductivity, plays a central role in determining thermoelectric performance [2–5]. In particular, the strategy known as band convergence [6–9], where multiple bands contribute to charge transport near the chemical potential $\mu$, serves as one of the most practical strategies to achieve high $zT$. Several high-performance materials have been successfully developed based on this concept, including PbTe-based systems with $zT \approx 2.5$ [10,11] and SnSe-based systems with $zT$ approaching 3.0 [12–14]. Nevertheless, quantitative design guidelines for band converged systems remain insufficiently established. For example, it remains unclear how close multiple band edges should be aligned to maximize $zT$, what band gap is required to suppress bipolar transport, and how the optimal chemical potential depends on the lattice thermal conductivity. Clarifying these issues is essential for translating the concept of band convergence into quantitative guidelines for thermoelectric materials design.

A major difficulty is that the influence of an individual band parameter is challenging to isolate in both first-principles calculations and experiments. In first-principles calculations for realistic materials, a change in a given band parameter is generally accompanied by changes in other band parameters, such as the band dispersion, effective mass, and band gap. In experiments, $\Delta E$ is often controlled by elemental doping [6,10,15], which can alter the band structure and carrier concentration while also modifying the relaxation time through impurity/alloy scattering. Therefore, it is challenging to reveal the intrinsic role of band convergence from either first-principles calculations or conventional transport experiments alone. To address this issue, a complementary and conceptually transparent approach is to employ simplified model calculations [4,5,16–20]. In such calculations, selected band parameters, such as the band offset $\Delta E$, can be varied independently while the other parameters are kept fixed. This enables us to isolate the effect of a selected band parameter on the thermoelectric properties. Despite its simplicity, such a model allows a systematic exploration of band structures with a high degree of flexibility, and it is useful for identifying general design principles for high thermoelectric performance [5,18,19].

In this study, we employ Boltzmann transport theory [5] for two-parabolic-band systems with independently tunable band parameters. This approach allows us to identify the band-structure conditions that maximize thermoelectric performance in band-converged systems. The resulting model provides quantitative and broadly applicable design guidelines for the development of thermoelectric materials.

## 2. Results

### 2.1 Energy window that affects $zT$

Using the two parabolic band model described in the Methods section, we calculated the thermoelectric properties as functions of the chemical potential. The transport coefficients were calculated within the framework of the Boltzmann transport theory [5] under the constant relaxation time approximation [21]. Our model assumes the $L$ band (light band) and $\Sigma$ band (heavy band) of PbTe for the two bands in the valence band, and the $L$ band (light band) for the conduction band [9,22]. Here, all the calculations were performed assuming $T = 900$ K, where $zT$ is maximized in most PbTe-based systems [10,11,15,23]. The spectral conductivity in the valence band is calculated as

$$\Sigma_i = \frac{e^2}{3} v_i^2(E)\tau_i(E)g_i(E) = \alpha \frac{\tau_i}{\tau_{L,v}} \frac{N_i}{N_{L,v}} \sqrt{\frac{m^*_{DOS,i}}{m^*_{DOS,L,v}}} \left(E_{i,v} - E\right)^{\frac{3}{2}} \Theta\left(E_{i,v} - E\right) \quad (1)$$

where $E$, $e$, $v_i$, $\tau_i$, and $g_i$ represent the carrier energy, elementary charge, group velocity, relaxation time, and the density of states (DOS) of band $i$, respectively. Here, the spectral conductivity is normalized by that of the $L$ band in the valence band. In the last equation, a three-dimensional isotropic parabolic band with a constant relaxation time is assumed (for details, see Supplementary Note 1 and Supplementary Fig. 1). The band parameters $\alpha$, $N_i$, $m^*_{DOS,i}$, and $E_{i,v}$ represent the normalization factor, band degeneracy, DOS effective mass, and band-edge energy of band $i$, respectively. The prefactor $\alpha$ was normalized to match the experimental value of electrical conductivity $\sigma$ of Sr/Na-codoped PbTe at $T = 900$ K [10], and we assumed that the total electrical conductivity of PbTe at $T = 900$ K is dominated by the $\Sigma$ band in the valence band [6] (for details, see Supplementary Note 3 and Supplementary Fig. 3). Although the $L$ band in PbTe is often approximated by the Kane model [24,25], we note that a parabolic-band approximation can also yield reasonable thermoelectric properties, as discussed in Supplementary

Note 1. The lattice thermal conductivity $\kappa_{lat}$ was set to constant values. Further details as well as band parameters are provided in the Methods section.

First, the DOS effective mass ratio dependence of $zT$ is investigated with a fixed band offset value $\Delta E = E_H - E_L = -0.5$ eV as schematically shown in Fig. 1(a). Here, $m^*_{DOS,H}$, $m^*_{DOS,L}$, $E_H$, and $E_L$ are the DOS effective mass [9] of the heavy band, that of the light band, band-edge energy of the heavy band, and that of the light band, respectively. Here, the origin of carrier energy is $E_L$, and this is the same for the chemical potential $\mu$. In this calculation, the DOS effective mass of the light band is fixed, and that of the heavy band is changed. Figure 1(b) shows the energy dependence of the spectral conductivity calculated for different DOS effective mass ratios $r_m = m^*_{DOS,H}/m^*_{DOS,L}$. Figures 1(c), (d), (e), and (f) present the chemical potential dependence of the figure of merit $zT$, electrical conductivity $\sigma$, Seebeck coefficient $S$, and electronic thermal conductivity $\kappa_{el}$, respectively. Here, the lattice thermal conductivity is fixed at $\kappa_{lat} = 0.5\ \mathrm{Wm^{-1}K^{-1}}$, a value reported in Sr/Na-codoped PbTe at $T = 900$ K [10]. As shown in Fig. 1(c), the maximum $zT$ is obtained at $\mu_{MAX} \approx 0.0$ eV, which is close to the band edge of the light band. Under this condition, the $r_m$ dependence of $zT$ is negligibly small; that is, the effective mass of the heavy band has little effect on $zT$. The same calculations were performed for a reduced band offset of $\Delta E = -0.3$ eV as shown in Fig. 2(a), and the calculated thermoelectric properties are shown in Figs. 2(b-f). In this case, the $zT$ values increase with increasing $r_m$ as shown in Fig. 2(c).

This behavior can be understood from the thermal window in the Boltzmann transport theory [5,26]. In this framework, the transport coefficients are expressed in terms of the transport integrals $L_n$, defined as

$$L_n = \int_{-\infty}^{\infty} (E-\mu)^n \Sigma(E) \left( -\frac{\partial f_{FD}(E,T)}{\partial E} \right) dE, \tag{2}$$

where $\mu$ is the chemical potential, and $f_{FD}(E,T)$ is the Fermi-Dirac distribution function. Using this expression, the electrical conductivity $\sigma$, Seebeck coefficient $S$, and electronic thermal conductivity $\kappa_{el}$ are given by [5]

$$\sigma = L_0, \tag{3}$$

$$S = -\frac{1}{eT}\frac{L_1}{L_0}, \tag{4}$$

$$\kappa_{el} = \frac{1}{e^2 T}\left( L_2 - \frac{L_1^2}{L_0} \right). \tag{5}$$

These expressions indicate that each transport coefficient is obtained by multiplying the spectral conductivity $\Sigma(E)$ by a window function and integrating over energy. The window functions for $\sigma, S,$ and the first term in $\kappa_{el}$ are given by

$$f_0 = \left(-\frac{\partial f_{FD}(E,T)}{\partial E}\right), \tag{6}$$

$$f_1 = (E-\mu)\left(-\frac{\partial f_{FD}(E,T)}{\partial E}\right), \tag{7}$$

$$f_2 = (E-\mu)^2\left(-\frac{\partial f_{FD}(E,T)}{\partial E}\right), \tag{8}$$

respectively. The energy dependence of these functions is shown in Figs. 3(a-c). As shown in Figs. 3(a-c), as $n$ increases, the effective energy window of the $f_n$ becomes wider. Specifically, the function $|f_1|$ decreases to approximately 10% of its maximum value at $(E-\mu)/k_BT \approx \pm 5.5$, where $k_B$ is the Boltzmann constant. This indicates that electronic states satisfying $|E-\mu| \geq 5k_BT$ contribute negligibly to thermoelectric transport in this case. Indeed, as seen in Fig. 1(c), the magnitude of the band offset $|\Delta E| = 0.5\ \mathrm{eV}$ is sufficiently larger than the thermal energy window of the Seebeck coefficient ($\approx 5k_BT \approx 0.39\ \mathrm{eV}$ at $T = 900\ \mathrm{K}$). As a result, the contribution of the heavy band to the transport coefficients is exponentially suppressed near $\mu = 0\ \mathrm{eV}$ and thermoelectric performance is governed almost entirely by the light band, leading to a negligible $r_m$ dependence of $zT$ in Fig. 1(c). In contrast, when $\Delta E$ is -0.3 eV, the heavy band lies within the $5k_BT$ window. Under this condition, increasing $r_m$ enhances the spectral conductivity near the chemical potential, leading to simultaneous increases in both the Seebeck coefficient $S$ and the electrical conductivity $\sigma$. Consequently, $zT$ near $\mu = 0\ \mathrm{eV}$ increases monotonically with $r_m$. We note that, within the present model, increasing $r_m$ effectively increases the magnitude of the spectral conductivity as can be seen from Eq. (1). Thus, high values of the relaxation time $\tau$ and the band degeneracy $N$ will similarly increase $zT$ in this model. Overall, achieving high spectral conductivity is crucial for obtaining high thermoelectric performance.

We emphasize that considering the thermal energy window shown in Fig. 3 is critically important for the design of high-performance thermoelectric materials, as pointed out by several previous studies [26,27]. Using the same framework, we can also evaluate the reduction of $zT$ due to the bipolar effect. To this end, we consider a system consisting of both valence and conduction bands as shown in Fig. 4(a), and perform calculations analogous to those shown in Fig. 1 and Fig. 2. The corresponding results are represented in Figs. 4(b-d). In these calculations, $\tau$, $N$, and $m_{DOS}{}^{*}$ are assumed to be identical for the conduction and valence bands [28,29]. As shown in Fig. 4(b), when $E_g \approx 5k_BT$ (corresponding to $E_g = 0.4$ eV in Fig. 4(b)), the reduction of $zT$ due to bipolar transport is limited to about 13%. This value is comparable to the typical experimental uncertainty (10-20% [30]) when evaluating $zT$ by conventional methods. To further investigate the bipolar effect on the individual thermoelectric properties, the $E_g$ dependences of $S$ and $\kappa_{el}$ are calculated as shown in Figs. 4(c) and 4(d), respectively. As shown in the insets in Figs. 4(c) and 4(d), the bipolar effect near $\mu = 0$ eV (band edge) is sufficiently suppressed when $E_g \geq 5k_BT$ for $S$ and $E_g \geq 7k_BT$ for $\kappa_{el}$. The $E_g$ dependences of $\sigma, S$, and $\kappa_{el}$ at $\mu = 0$ eV, which are normalized to their respective values at $E_g = 10k_BT$ and $\mu = 0$, are summarized in Fig. 4(e). To define a practical criterion for suppressing the bipolar effect, we set the thresholds such that its impact on $zT$ does not exceed approximately 10%. Accordingly, we set the allowable bipolar-induced reduction in $|S|$ to 5%, because $S$ enters the expression for $zT$ quadratically, whereas we set the allowable bipolar-induced increase in $\kappa_{el}$ to 10%. Following these criteria, we estimate the minimum band gap required for thermoelectric materials to be $E_g \geq 5k_BT$ for $S$ and $E_g \geq 7k_BT$ for $\kappa_{el}$.

To experimentally examine the validity of the $5k_BT$ criterion for $S$, we further analyzed the experimental data on band gap and temperature dependence of $S$ for representative thermoelectric materials [10,31–49]. For each material system, we selected the composition exhibiting the highest thermoelectric performance among those reported in the corresponding study. We then defined $T_{BP}$ as the temperature at which $|S|$ begins to decrease with increasing temperature; when no decrease was observed, the highest measured temperature was taken as a lower bound for $T_{BP}$ and indicated by an arrow in Fig. 4(f). As shown in Fig. 4(f), $T_{BP}$ increases approximately linearly with $E_g$, and the experimental data broadly follow the relation $T_{BP} = E_g/5k_B$. This trend provides experimental support for the $5k_BT$ criterion obtained from our model calculations. We attribute this behavior to the fact that, near the optimal carrier concentration, the chemical potential typically lies close to the band edge [50]. Consequently, bipolar transport begins to become appreciable when the energy separation between the chemical potential and the minority-carrier band edge becomes approximately $5k_BT$. The experimentally observed relation $E_g \approx 5k_BT_{BP}$ also suggests that the band gap can be estimated from the temperature dependence of the Seebeck coefficient for samples with near-optimal carrier concentrations.

Lastly, we discuss the applicability of these criteria. Within the constant relaxation time approximation, the spectral conductivity of a parabolic band is expressed as $\Sigma(E) = \Sigma_0 \varepsilon^{\beta}$ with $\beta = 1.5$, where $\varepsilon$ is the energy measured from the relevant band edge and $\Sigma_0$ is an energy-independent prefactor. Since $\kappa_{el}$ is proportional to $\Sigma_0$, the ratio $\kappa_{el}(E_g = 7k_BT)/\kappa_{el}(E_g = 10k_BT)$ is independent of $\Sigma_0$. A similar argument also applies to $\sigma$ and $S$. In Supplementary Information, we additionally performed similar calculations for two additional $\beta$ values ($\beta = 1.0$ and $2.0$). The $5k_BT$ and $7k_BT$ criteria nevertheless appear robust, possibly due to the exponential decay of the window functions in Fig. 3 at $|E-\mu|/k_BT > 5$ or $|E-\mu|/k_BT > 7$. We believe that the $5k_BT$ criterion for $S$ and the $7k_BT$ criterion for $\kappa_{el}$ can serve as quantitative guidelines for the band gap required for thermoelectric materials.

## 2.2 Enhancement of $zT$ by band convergence

To investigate the effect of band convergence on $zT$ values, we consider two parabolic bands in the valence band with $m^*_{DOS,H}/m^*_{DOS,L} = 5$, as schematically shown in Fig. 5(a). Here, we used the literature values indicating that the ratio of the DOS effective mass between the $\Sigma$ band and $L$ band in PbTe is $m^*_{\mathrm{DOS},\Sigma}/m^*_{DOS,L} \approx 5$ [9]. Detailed descriptions of band parameters are provided in the Methods section. The resulting thermoelectric properties are shown in Figs. 5(b-d). Figure 5(b) presents $zT$ as a function of the chemical potential for various values of the band-edge energy separation $\Delta E$. When $\Delta E = -0.5$ eV, corresponding to $|\Delta E| > 5k_BT$, the heavy band lies far below the light band and the $zT$ profile coincides with that of the light band (see Fig. 1(c)). As shown in the inset of Fig. 5(d), when $\Delta E$ is increased from $-0.5$ eV toward $0$ eV, the Seebeck coefficient near $\mu = 0.0$ eV first increases and then decreases. This behavior can be understood as follows. Explicitly, the Seebeck coefficient in Eq. (4) can be rewritten as:

$$S = -\frac{1}{eT}\frac{\int_{-\infty}^{\infty}\Sigma(E)\,f_1 dE}{\int_{-\infty}^{\infty}\Sigma(E)f_0\,dE} = -\frac{1}{eT}\frac{\int_{-\infty}^{\infty}\Sigma(E)f_1\,dE}{\sigma}. \quad (9)$$

When $\Delta E$ increases from $-0.5$ eV, the Seebeck coefficient $S$ initially increases because the $5k_BT$ energy window associated with the $f_1$ function is larger than that of the $f_0$ function (see Fig. 3(a) and 3(b)). However, when $\Delta E$ approaches $0$ eV, the electrical conductivity $\sigma$ increases substantially, leading to a decrease in $S$ (see Eq. (9)).

As shown in Fig. 5(b), the maximum $zT$ is attained when the band edges of the two bands are completely aligned ($\Delta E = 0$: gray curve), consistent with previous first-principles studies based on the constant relaxation time approximation [51,52]. The inset of Fig. 5(d) indicates that, at the optimized chemical potential $\mu \approx 0.092$ eV ($zT$ is maximized for $\Delta E = 0$), the enhancement of the Seebeck coefficient by band convergence is negligible, with a ratio of $S(\Delta E = 0)/S(\Delta E = -0.5\text{ eV}) \approx 1.0$ (see the inset of Fig. 5(d)). In contrast, a substantial enhancement of the electrical conductivity $\sigma$ is observed, as evidenced by the ratio $\sigma(\Delta E = 0)/\sigma(\Delta E = -0.5\text{ eV}) \approx 3.5$, as shown in the inset of Fig. 5(c). The electronic thermal conductivity $\kappa_{el}$ also increases, with $\kappa_{el}(\Delta E = 0)/\kappa_{el}(\Delta E = -0.5\text{ eV}) \approx 3.4$, as shown in Supplementary Information. However, because the total thermal conductivity is given by $\kappa = \kappa_{el} + \kappa_{lat}$, the enhancement of $\kappa$ is limited to $\kappa(\Delta E = 0)/\kappa(\Delta E = -0.5\text{ eV}) \approx 1.5$, which ultimately leads to an enhancement of $zT$.

We note that the discussion above assumes a fixed chemical potential. At fixed carrier concentration, on the other hand, band convergence shifts the chemical potential toward the band gap, which can equivalently be viewed as the origin of the enhanced Seebeck coefficient. We also observe from Fig. 5(b) that while the maximum $zT$ values significantly decrease for $\Delta E < 0$, the reduction in the maximum $zT$ is much smaller for $\Delta E > 0$ (when the heavy band dominates the transport properties). These results also indicate that large values of the spectral conductivity $\Sigma(E)$ near the band edge are essentially the most important factor for achieving high $zT$ values. We note that the predicted $zT$ in Fig. 5(b) is reasonably consistent with the highest value reported in experiments ($zT \approx 2.5$ in Sr/Na codoped PbTe [10,11]), although no additional fitting parameters were introduced beyond the normalization procedure described in the Methods section. This point validates the present transport model including the normalization process.

To experimentally validate band convergence, a Jonker plot is a useful approach [17,53]. Under the approximation of non-degenerate semiconductors and constant relaxation time, the Seebeck coefficient can be expressed as

$$S = \frac{k_B}{e}\left[\frac{5}{2} - \ln\sigma + \ln\left\{2Ne^2 m^{*\frac{1}{2}}_{DOS}\tau\left(\frac{k_B T}{2\pi\hbar^2}\right)^{\frac{3}{2}}\right\}\right]. \quad (10)$$

Thus, plotting $S$ as a function of $\ln\sigma$ for a given material with different carrier concentrations should yield a linear relation. Since the Seebeck coefficient $S$ and electrical conductivity $\sigma$ have been calculated from the Boltzmann transport theory in Figs. 5(c) and 5(d), the results can be directly compared with those predicted by the Jonker relation in Eq. (10). Figure 5(e) shows the results obtained from the Boltzmann transport theory (curves), together with experimental data for $Pb_{0.98}Na_{0.02}Te$-x%SrTe [10] and $Pb_{1-x}Na_xTe$ [15]. At a fixed $\sigma$, the curves for $\Delta E \approx 0$ exhibit larger values of the Seebeck coefficient as shown in Fig. 5(e). This enhancement originates from the enhanced effective band degeneracy $N$ arising from band convergence, and such behavior is indeed observed in the experimental data [10,15,23]. Figure 5(e) shows that Sr, Na and Mg doping in PbTe bring the data points closer to the $\Delta E = 0$ curves (gray). This behavior is consistent with previous first-principles calculations showing that Sr and Na doping induces band convergence [10,15]. In addition to enhancing the $N$ value (band convergence), Eq. (10) indicates that the enhancement of the DOS effective mass $m_{DOS}^*$ and the relaxation time $\tau$ would further enhance the $zT$ values. These new strategies will serve as useful guidelines for achieving even higher $zT$ values in thermoelectric materials.

To further investigate band convergence, a Pisarenko plot is performed for the calculation results from the Boltzmann transport theory, together with experimental data [10,15,23]. Under the approximation of parabolic band and constant relaxation time approximation, the Seebeck coefficient can be expressed as

$$S = \frac{k_B}{e}\left[\frac{5}{2} - \ln n + \ln\left\{2N\left(\frac{m_{DOS}^{*}k_B T}{2\pi\hbar^2}\right)^{\frac{3}{2}}\right\}\right]. \quad (11)$$

Since Eq. (11) does not contain $\tau$, a Pisarenko plot is useful for probing changes in $m_{DOS}^{*}$. As shown in Fig. 5(f), the increase in $m_{DOS}^{*}$ upon Na [15], Sr [10] and Mg [23] doping can be demonstrated experimentally. However, as also shown in Fig. 5(f), it is difficult to experimentally establish whether $\Delta E = 0$ from the Pisarenko plot alone. Therefore, spectroscopic measurements [22,33,54] are required for such a verification.

### 2.3 Optimal chemical potential

Unlike the absolute value of the spectral conductivity or the band offset $\Delta E$, it is not possible to specify a universal optimal position of the chemical potential $\mu$ that maximizes $zT$. This is because the optimal $\mu$ strongly depends on the relative magnitude of the electronic thermal conductivity $\kappa_{el}$ and the lattice thermal conductivity $\kappa_{lat}$. We discuss this point in detail below. In the limit of $\kappa_{lat} \gg \kappa_{el}$, maximizing $zT$ essentially reduces to maximizing the power factor $\sigma S^2$. Within the constant relaxation time approximation for a three-dimensional single parabolic band, the optimal chemical potential can be calculated to be located at $\mu \approx E_{edge} - 2.5k_B T$ (for details, see Supplementary Note 2 and Supplementary Fig. 2), indicating that the chemical potential should lie within the band. In contrast, in the opposite limit ($\kappa_{el} \gg \kappa_{lat} = 0$), the figure of merit can be written as $zT = \sigma S^2 T/\kappa_{el} = S^2/L$, where the Wiedemann-Franz law has been used in the last equation and $L$ is the Lorenz number. Under this condition, $zT$ increases monotonically as the chemical potential moves deeper into the band gap, implying that the optimal $\mu$ would be located at the middle of the band gap. However, this conclusion leads to an apparent contradiction when a finite lattice thermal conductivity $\kappa_{lat}$ is taken into account. When the chemical potential is placed near the center of the band gap, the electronic thermal conductivity $\kappa_{el}$ becomes sufficiently small that $\kappa_{el}$ is no longer larger than $\kappa_{lat}$. Consequently, the assumption $\kappa_{el} \gg \kappa_{lat}$ breaks down. Thus, the lattice thermal conductivity must be explicitly taken into account when determining the optimal chemical potential for maximizing $zT$.

We next consider the two-valence-band system shown in Fig. 6(a), whose band parameters are identical to those in Fig. 5(a), while varying the lattice thermal conductivity $\kappa_{lat}$. We calculate the chemical potential dependence of $zT$ with several $\kappa_{lat}$ values: $\kappa_{lat} = 0\ \mathrm{Wm^{-1}K^{-1}}$ in Fig. 6(b), $\kappa_{lat} = 0.25\ \mathrm{Wm^{-1}K^{-1}}$ in Fig. 6(c), $\kappa_{lat} = 0.5\ \mathrm{Wm^{-1}K^{-1}}$ in Fig. 6(d), and $\kappa_{lat} = 10\ \mathrm{Wm^{-1}K^{-1}}$ in Fig. 6(e). Here, $\kappa_{lat} = 0.25\ \mathrm{Wm^{-1}K^{-1}}$ has been reported in SnSe-based systems [14], and $\kappa_{lat} = 0.5\ \mathrm{Wm^{-1}K^{-1}}$ in PbTe-based systems [10]. As expected, the optimal chemical potential is located at the middle of the band gap when $\kappa_{lat} = 0\ \mathrm{Wm^{-1}K^{-1}}$ as shown in Fig. 6(b). When $\kappa_{lat} = 0.25\ \mathrm{Wm^{-1}K^{-1}}$, the optimal chemical potential is $\mu \approx +0.13$ eV above the band edge, and it is $\mu \approx +0.092$ eV for $\kappa_{lat} = 0.5\ \mathrm{Wm^{-1}K^{-1}}$. In contrast, when the lattice contribution dominates the total thermal conductivity ($\kappa_{lat} = 10\ \mathrm{Wm^{-1}K^{-1}}$), the optimal chemical potential is located below the band edge at $\mu \approx -0.065$ eV. The relation between the lattice thermal conductivity $\kappa_{lat}$ and optimal chemical potential $\mu_{opt}$ for $\Delta E = 0$ is summarized in Fig. 6(f), together with additional data points. These results demonstrate that, although a universal optimal value of $\mu_{opt}$ does not exist, a clear trend emerges. Specifically, when the lattice thermal conductivity $\kappa_{lat}$ dominates the total thermal conductivity, $\mu_{opt}$ approaches $-2.5k_BT$, where the chemical potential crosses the band edge, as discussed in the previous paragraph. In contrast, as $\kappa_{lat}$ decreases, the optimal chemical potential $\mu_{opt}$ shifts progressively away from the band edge toward the center of the band gap, corresponding to smaller carrier concentrations.

### 2.4 Modifications of our model

In the present study, we employ a simplified transport model to identify optimal band parameters in band-converged systems. Several effects were deliberately neglected, including the energy dependence of the relaxation time [55,56] and interband scattering [57,58].

Since first-principles calculations on PbTe indicate that the energy dependence of the relaxation time for longitudinal optical phonon modes is relatively small [9] ($\tau(E) = \tau_0 E^\gamma$ with $\gamma \approx -0.2$ for the $\Sigma$ band), the energy dependence of the relaxation time is neglected in the main text of the present study. Similar calculations can be performed with an energy-dependent relaxation time; however, the qualitative trends remain the same (see Supplementary Note 4 and Supplementary Fig. 4). The energy dependence of the relaxation time has often been approximated as a simple power-law: $\tau(E) = \tau_0 E^\gamma$. The values of $\gamma$ are $-1/2, 1/2,$ and $3/2$ for acoustic phonon scattering, polar optical phonon scattering, and ionized impurity scattering, respectively. Since all calculations in this work are performed at $T = 900$ K, ionized impurity scattering is expected to be negligible. In Supplementary Note 4, we performed similar calculations to those shown in Figs. 2 and 5 for $\gamma = -1/2$ and $1/2$. Although the absolute values of $zT$ are enhanced in the case of $\gamma = 1/2$ when compared with $\gamma = 0$ in the main text, the overall trends remain unchanged: $zT$ is enhanced when the absolute value of the spectral conductivity is large, and when the band offset $\Delta E$ is zero.

Interband scattering was also neglected in the present calculations. In realistic materials, strong interband scattering reduces carrier mobility and, in certain cases, could suppress the power factor despite the enhancement of the DOS by band convergence [57]. In light of the present results, if band convergence leads to smaller spectral conductivity than that of the single band cases, $zT$ is expected to decrease upon band convergence. In this case, the optimal condition may shift from $\Delta E = 0$ to a finite positive $\Delta E$, where the band with larger spectral conductivity dominates the transport properties. Nevertheless, even in such cases, the essential requirement for achieving high $zT$ remains the enhancement of the spectral conductivity near the band edge; that is, high band degeneracy $N$, DOS effective mass $m_{DOS}^*$, and relaxation time $\tau$ are required.

## 3. Conclusion

In this study, by employing Boltzmann transport theory, we have identified a set of band-structure parameters that are favorable for achieving high thermoelectric performance. The main conclusions of this study are as follows: (i) To suppress the bipolar effect, the band gap $E_g \geq 5k_BT$ is required for $S$ and $E_g \geq 7k_BT$ for $\kappa_{el}$. The $5k_BT$ criterion for $S$ is also consistent with experimental data for representative thermoelectric materials, which broadly follow $T_{BP} \approx$ $E_g/(5k_B)$. (ii) Band convergence with $\Delta E$ close to zero provides optimal thermoelectric performance when interband scattering is insignificant. (iii) The optimal chemical potential is determined by the relative magnitude of $\kappa_{el}$ and $\kappa_{lat}$, and shifts toward the band gap as $\kappa_{lat}$ decreases. (iv) Large values of the spectral conductivity (high band degeneracy $N$, DOS effective mass $m^*_{DOS}$, and relaxation time $\tau$) are crucial for enhancing $zT$. The present work establishes quantitative material-design guidelines for band-converged thermoelectric materials, providing a practical framework for the rational design of high-performance thermoelectric materials.

## 4. Methods

### 4.1 Model framework and transport formalism

Thermoelectric transport properties were calculated within the framework of the Boltzmann transport theory [3,5] under the constant relaxation time approximation. The transport coefficients were expressed in terms of the transport integrals $L_n$ as shown in Eq. (2). Using this expression, the electrical conductivity $\sigma$, Seebeck coefficient $S$, and electronic thermal conductivity $\kappa_{el}$ were calculated by Eqs. (3-5). The dimensionless figure of merit was then evaluated as

$$zT = \frac{\sigma S^2 T}{\kappa_{el} + \kappa_{lat}}, \tag{12}$$

where the lattice thermal conductivity $\kappa_{lat}$ was set to constant values.

### 4.2 Two-band spectral conductivity model in the valence band

To investigate the effect of band convergence, a simplified two-parabolic-band model was employed. Within the constant relaxation time approximation, the spectral conductivity in the valence band was calculated as the sum of contributions from two parabolic bands,

$$\Sigma(E) = \sum_{i=L,H} \Sigma_i = \sum_{i=L,H} \alpha \frac{\tau_i}{\tau_{L,v}} \frac{N_i}{N_{L,v}} \sqrt{\frac{m^*_{DOS,i}}{m^*_{DOS,L,v}}} \left(E_{i,v} - E\right)^{\frac{3}{2}} \Theta\left(E_{i,v} - E\right), \tag{13}$$

where $\alpha$, $\tau_i$, $N_i$, $m^*_{DOS,i}$, and $E_{i,v}$ represent the normalization factor, relaxation time, band degeneracy, DOS effective mass, and the band-edge energy of band $i$, respectively. Our model assumes the $L$ band (light band) and $\Sigma$ band (heavy band) of PbTe for the two bands in the valence band, and the $L$ band (light band) for the conduction band [9,22]. Here, the spectral conductivity is normalized by that of the $L$ band in the valence band. The band parameters in Eq. (13) are presented in Table 1. The exponent $3/2$ corresponds to a three-dimensional parabolic band under the constant relaxation time approximation (for details, see Supplementary Note 1). The Heaviside step function $\Theta(x)$ is introduced to explicitly enforce that only states with energies below the band edge ($E \leq E_{i,v}$) contribute to the spectral conductivity. The energy separation between the two band edges was defined as

$$\Delta E = E_{H,v} - E_{L,v}. \tag{14}$$

The bipolar effect was neglected in all calculations except for Fig. 4, as the band gap of pristine PbTe ($E_g \approx 0.48$ eV at $T = 900$ K) is sufficiently larger than $5k_BT$ ($\approx 0.39$ eV at $T = 900$ K) and it further increases by Sr doping [10,22].

### 4.3 Extension to conduction bands

For conduction-band transport, the spectral conductivity was defined analogously as

$$\Sigma_c = \alpha \frac{\tau_c}{\tau_{L,v}} \frac{N_c}{N_{L,v}} \sqrt{\frac{m^*_{DOS,c}}{m^*_{DOS,L,v}}} (E - E_c)^{\frac{3}{2}} \Theta(E - E_c), \tag{15}$$

where $E_c$ and $m^*_{DOS,c}$ denote the conduction-band edge energy and the corresponding DOS effective mass, respectively.

### 4.4 Normalization factor of spectral conductivity $\alpha$ and band parameters

To obtain a reasonable magnitude of the electrical conductivity, the spectral conductivity was normalized so that the electrical conductivity matches the experimental value at a reference chemical potential $\mu_{ref}$ and temperature $T_{ref}$. Specifically, a normalization factor $\alpha$ was calculated from the condition

$$\alpha = \frac{\sigma_{exp}}{L_0^{ref}(\alpha = 1)}, \tag{16}$$

where $L_0^{ref}$ is evaluated for a reference band at $\mu = \mu_{ref}$ and $T_{ref} = 900$ K assuming $\alpha = 1$.

The target electrical conductivity $\sigma_{exp}$ is set to $\sigma_{exp} = 300\ (\Omega\mathrm{cm})^{-1}$, which is reported for Sr/Na-doped PbTe at $T = 900\ \mathrm{K}$ in a previous experimental study [10]. At $T = 900\ \mathrm{K}$, we assume that the total electrical conduction is dominated by the valence band at the $\Sigma$ point ($\Sigma$ band) as shown in a previous study of Na-doped $PbTe_{1-x}Se_x$ [6]. Accordingly, the total carrier concentration $n$ is also primarily dominated by the $\Sigma$ band. The validity of this assumption is discussed in detail in Supplementary Note 3. Thus, the reference chemical potential $\mu_{ref}$ can be determined by solving the following equation for a given $n$.

$$n = \frac{\left(2m_{DOS,\Sigma}^{*}\right)^{\frac{3}{2}}}{2\pi^2\hbar^3} N_\Sigma \int_{-\infty}^{E_{\Sigma,v}} \frac{\left(E_{\Sigma,v} - E\right)^{\frac{1}{2}}}{1 + \exp\frac{\mu_{ref} - E}{k_B T}} dE, \tag{17}$$

where $m_{DOS,\Sigma}^{*}$, $N_\Sigma$, $E_{\Sigma,v}$,and $\hbar$ denote the DOS effective mass, band degeneracy, band-edge energy of the $\Sigma$ band, and reduced Planck constant, respectively. From our previous study, the upper limit of carrier concentration $n$ in Sr/Na-doped PbTe, which gives the highest $zT$ value, is estimated to be $n \approx 1.3 \times 10^{20}\ \mathrm{cm}^{-3}$ [22]. The results in Supplementary Note 5 and Supplementary Fig. 5 show that $n \approx 1.3 \times 10^{20}\ \mathrm{cm}^{-3}$ corresponds to $\mu = E_{\Sigma,v} + 60\ \mathrm{meV}$ at $T = 900\ \mathrm{K}$. Using this value of $\mu_{ref}$, the normalization factor $\alpha$ is determined by Eq. (16).

Correspondingly, $\Sigma(E)$ for the $L$ band can be calculated by Eq. (13) using the same $\alpha$ value and the band parameters listed in Table 1. Here, the ratio of the relaxation time $\tau_\Sigma/\tau_L$ was estimated using the mobility ratio $\mu_\Sigma/\mu_L = 0.075$ [22], which was experimentally determined by the two-carrier model in Sr/Na-codoped PbTe.

In the calculations of Figs. 1 and 2, $\alpha$ was first determined using the same procedure described above. The spectral conductivity was then calculated using Eq. (13) with different mass ratios $r_m = m^*_{DOS,\mathrm{H},v_}/m^*_{DOS,L,v}$.

**Credit author statement**

Y.H. conceived the study, performed the computations, and wrote the manuscript under the supervision of Y.M., who contributed to the study design and manuscript preparation. H.U. contributed to the interpretation of the results and manuscript writing.

**Declaration of competing interest**

The authors declare that they have no known competing financial interests or personal relationships that could have appeared to influence the work reported in this paper.

**Acknowledgements**

Y.H. acknowledges financial support by JSPS KAKENHI (Grant No. 24KJ0227) and H.U. by JSPS KAKENHI (Grant No. 24K08231). The authors also thank A. Novitskii and I. Serhiienko for fruitful discussions.

## Data availability

The authors declare that all data supporting the findings of this study are available within the article and its Supplementary Information files or from the corresponding author.

## Appendix A. Supplementary data

Supplementary data related to this article can be found online at ...

**Table 1.** Band parameters used in the present spectral conductivity model.

| | $\tau/\tau_{L,v}$ | $N/N_{L,v}$ | $m^*_{DOS}/m^*_{DOS,L,v}$ |
|---|---|---|---|
| Heavy band in valence band | 0.375 | $12/4 = 3$ | 5 |
| Light band in conduction band | 1 | 1 | 1 |
| Light band in valence band | 1 | 1 | 1 |

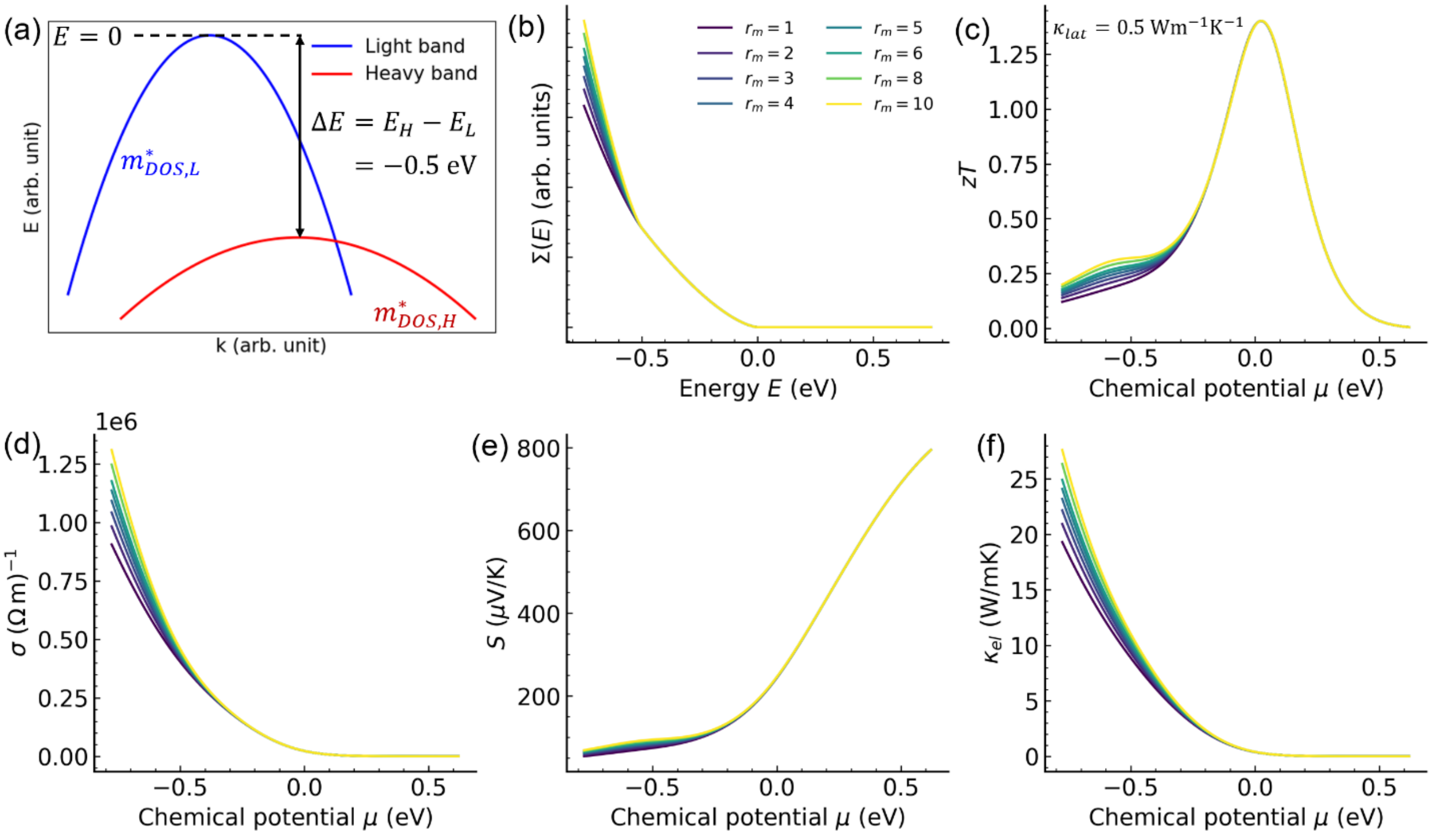

**Fig. 1 Thermoelectric properties calculated for two valence bands with an energy separation of $\Delta E = -0.5$ eV.** (a) Schematic illustration of two valence bands. (b) Corresponding spectral conductivity for different DOS effective mass ratio $r_m$, which is used to calculate the chemical

potential $\mu$ dependence of (c) the thermoelectric figure of merit $zT$, (d) electrical conductivity $\sigma$, (e) Seebeck coefficient $S$, and (f) electronic thermal conductivity $\kappa_{el}$. Here, the temperature is set to $T = 900$ K and the lattice thermal conductivity is $\kappa_{lat} = 0.5$ Wm$^{-1}$K$^{-1}$.

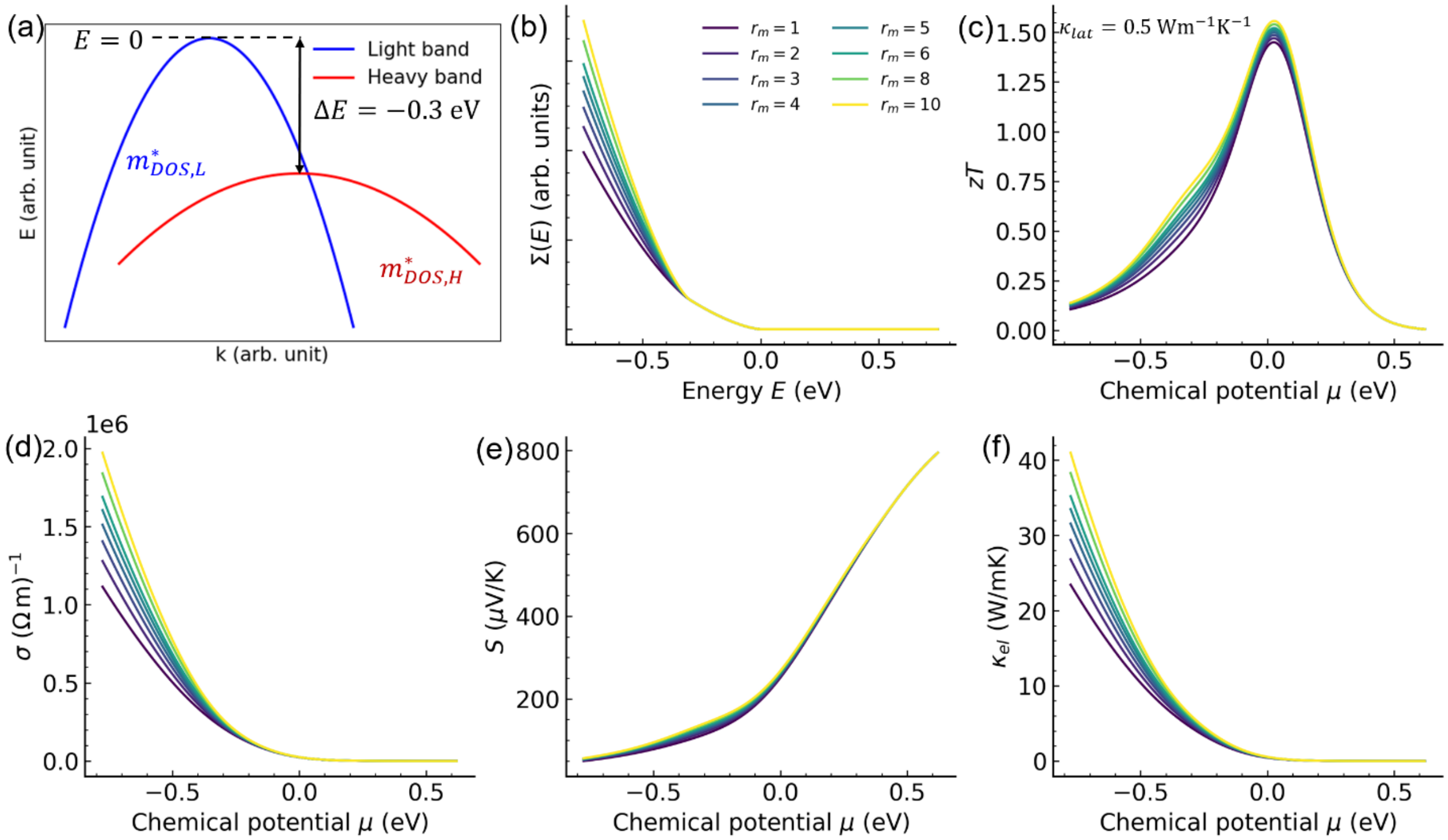

**Fig. 2 Thermoelectric properties calculated for two valence bands with an energy separation of $\boldsymbol{\Delta E = -0.3}$ eV.** (a) Schematic illustration of two valence bands. (b) Corresponding spectral conductivity for different DOS effective mass ratio $r_m$, which is used to calculate the chemical potential $\mu$ dependence of (c) the thermoelectric figure of merit $zT$, (d) electrical conductivity $\sigma$, (e) Seebeck coefficient $S$, and (f) electronic thermal conductivity $\kappa_{el}$. Here, the temperature is set to $T = 900$ K and the lattice thermal conductivity is $\kappa_{lat} = 0.5$ Wm$^{-1}$K$^{-1}$.

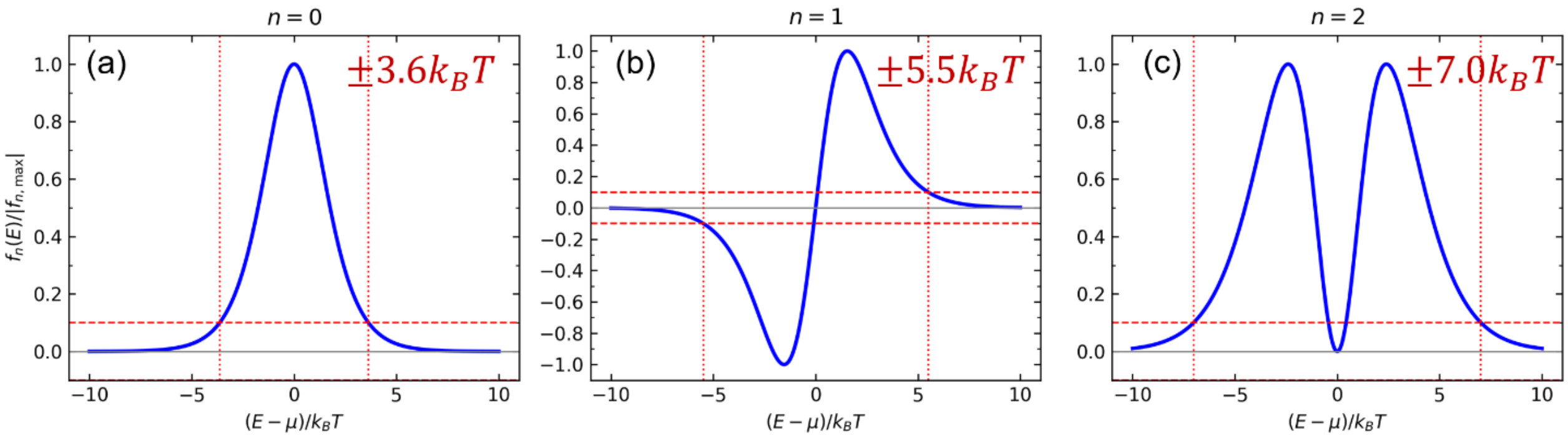

**Fig. 3 Energy dependence of the function $f_n = (E - \mu)^n(-\partial f_{FD}(E,T)/\partial E)$ for** (a) $n = 0$, (b) $n = 1$, and (c) $n = 2$. Here, $f_{FD}(E,T)$ is the Fermi-Dirac distribution function.

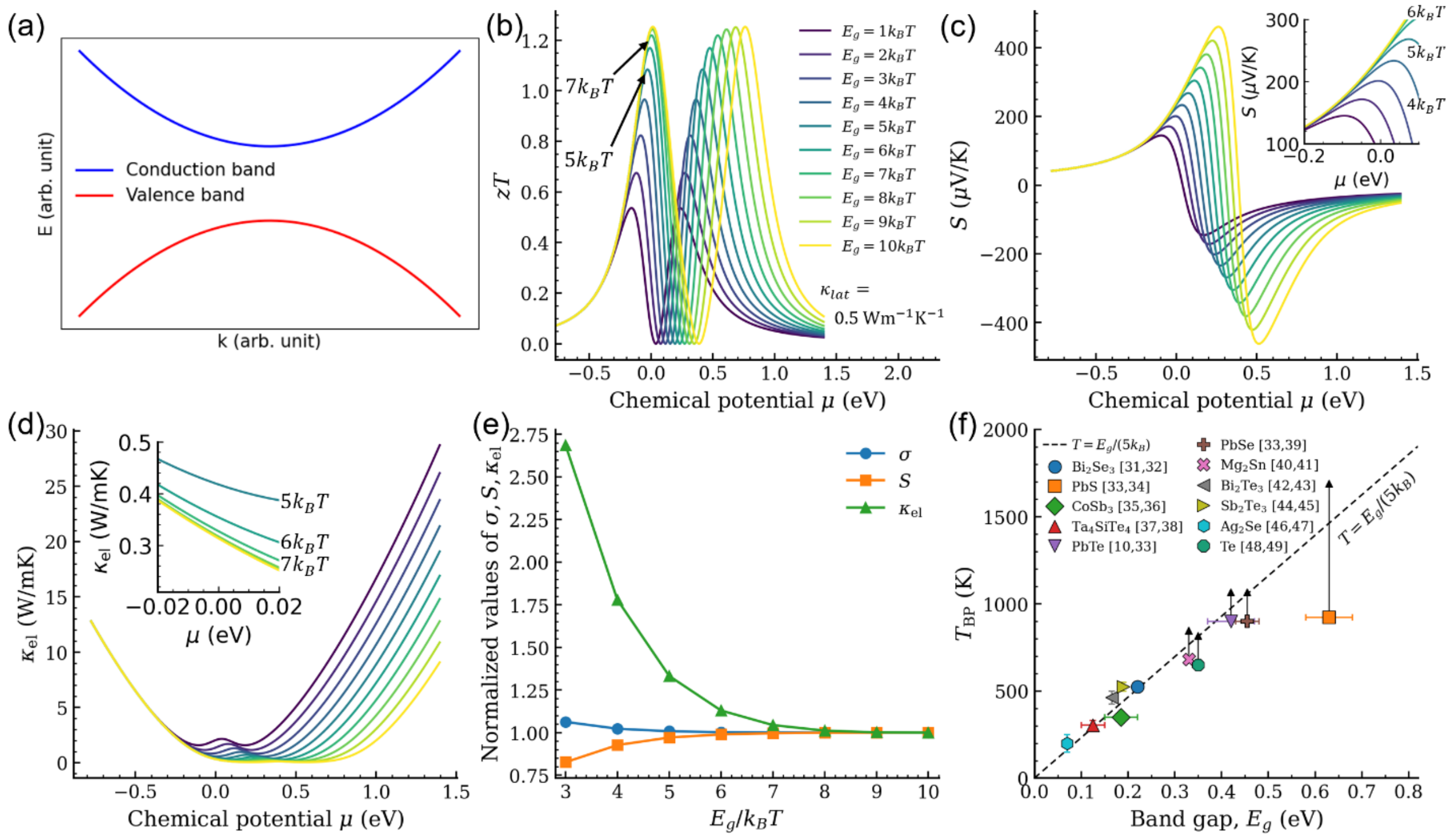

**Fig. 4 Thermoelectric properties calculated for a valence band and a conduction band for several band gaps $E_g$.** (a) Schematic illustration of a conduction band and a valence band. The chemical potential $\mu$ dependence of the calculated (b) thermoelectric figure of merit $zT$, (c) Seebeck coefficient $S$, and (d) electronic thermal conductivity $\kappa_{el}$. Here, the temperature is set to $T = 900$ K and the lattice thermal conductivity is $\kappa_{lat} = 0.5\ \mathrm{Wm^{-1}K^{-1}}$. (e) $E_g$ dependences of $\sigma, S, \kappa_{el}$, which are normalized by those respective values at $E_g = 10k_BT$ and $\mu = 0$. (f) Band gap dependence of $T_{BP}$, defined as the temperature at which $|S|$ begins to decrease with increasing temperature for samples with near-optimal carrier concentrations. For materials that do not exhibit the bipolar effect within the measured temperature range, the highest measured temperature is marked with an arrow.

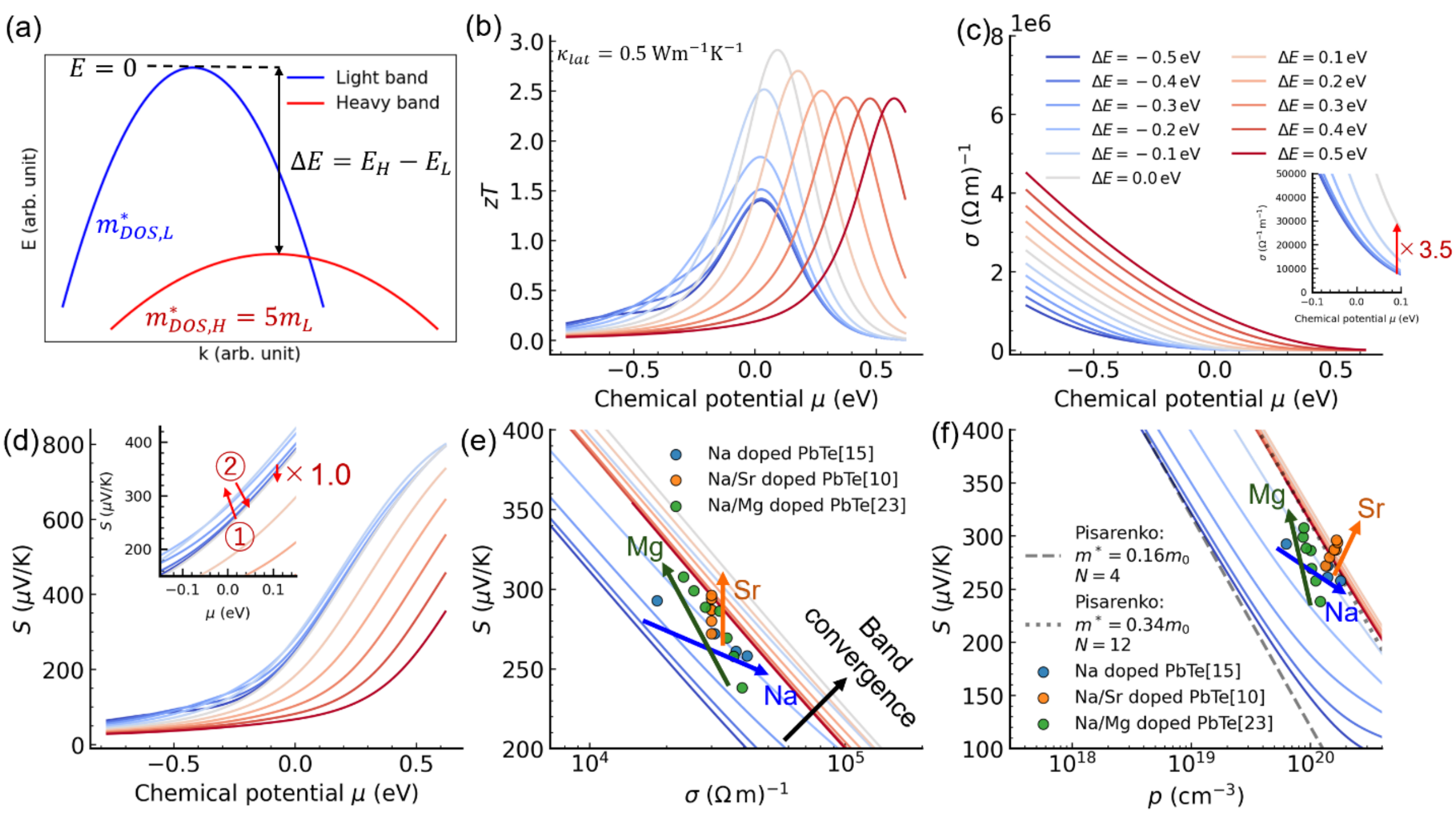

**Fig. 5 Thermoelectric properties calculated for two valence bands with several band offsets $\Delta E$.** (a) Schematic illustration of two valence bands. (b-d) Chemical potential $\mu$ dependence of (b) the thermoelectric figure of merit $zT$, (c) electrical conductivity $\sigma$, (d) Seebeck coefficient $S$. Insets in Figs. 5(c) and 5(d) show enlarged views. (e) Jonker plot and (f) Pisarenko plot for the calculation results from the Boltzmann transport theory, together with experimental data points for $Pb_{0.98}Na_{0.02}Te$-x%SrTe [10], $Pb_{1-x}Na_xTe$ [15] and $Pb_{0.98}Na_{0.02}Te$+x%MgTe [23]. Here, the temperature is set to $T = 900$ K and the lattice thermal conductivity is $\kappa_{lat} = 0.5\ \mathrm{Wm^{-1}K^{-1}}$.

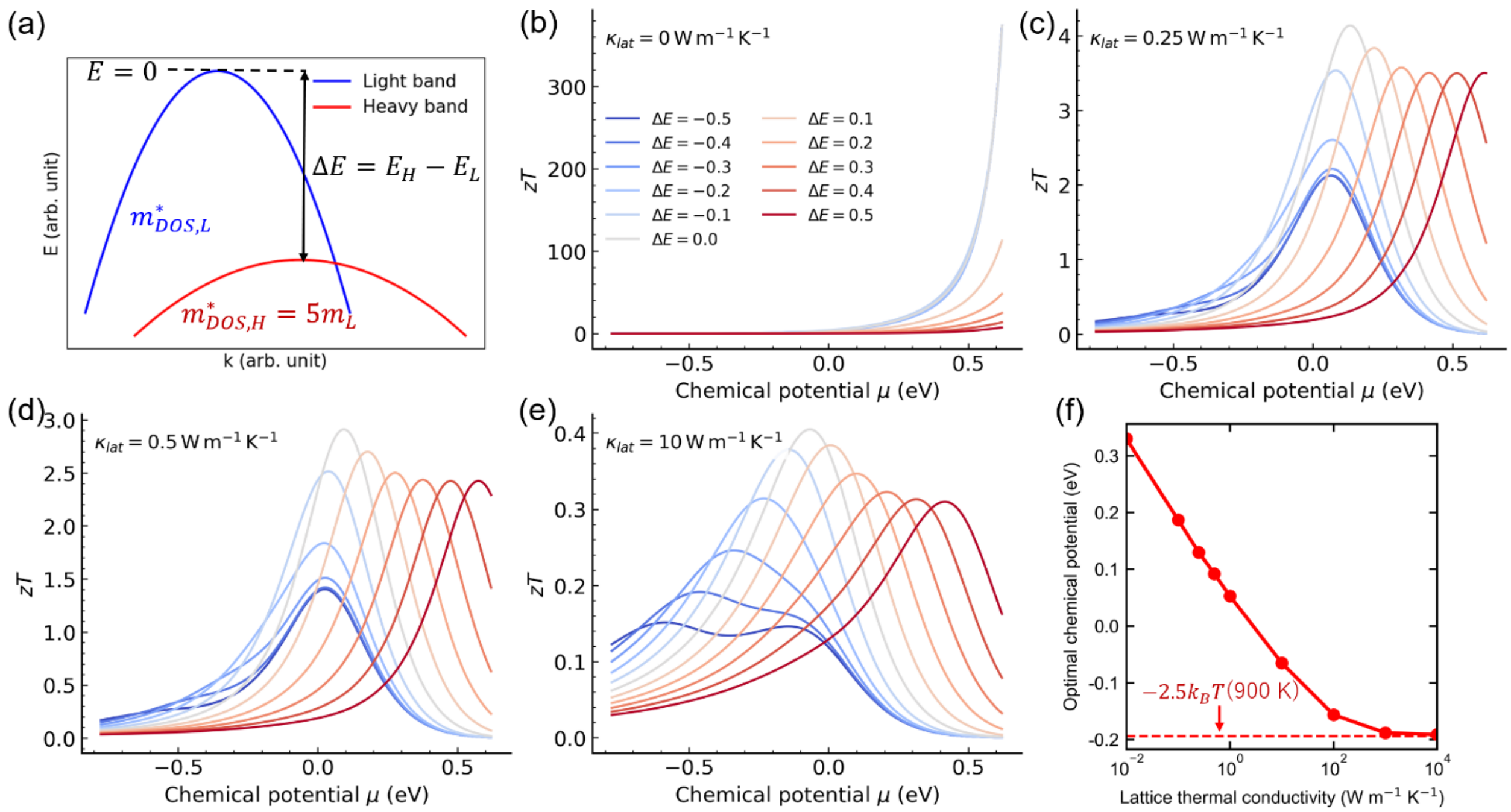

**Fig. 6 Thermoelectric properties of two valence bands calculated for several band offsets $\Delta E$ with different $\kappa_{lat}$ values.** (a) Schematic illustration of two valence bands. The chemical

potential $\mu$ dependence of the thermoelectric figure of merit $zT$ is calculated with (b) lattice thermal conductivity $\kappa_{lat} = 0\ \mathrm{Wm^{-1}K^{-1}}$, (c) lattice thermal conductivity $\kappa_{lat} = 0.25\ \mathrm{Wm^{-1}K^{-1}}$, (d) lattice thermal conductivity $\kappa_{lat} = 0.5\ \mathrm{Wm^{-1}K^{-1}}$, and (e) lattice thermal conductivity $\kappa_{lat} = 10\ \mathrm{Wm^{-1}K^{-1}}$. (f) Relationship between lattice thermal conductivity $\kappa_{lat}$ and the optimal chemical potential $\mu_{opt}$ for $\Delta E = 0$. The horizontal axis is shown on a logarithmic scale. Here, the temperature is set to $T = 900$ K.

# Supplementary Information

# Ideal band structures for high-performance thermoelectric materials with band convergence

Yuya Hattori[1,a], Hidetomo Usui[2] and Yoshikazu Mizuguchi[1]

*[1]Department of Physics, Tokyo Metropolitan University, Tokyo 192-0397, Japan*

*[2]Department of Engineering Science, The University of Electro-Communications, Tokyo 182-8585, Japan*

[a] Corresponding author's E-mail: yhattori@tmu.ac.jp

## Supplementary Note 1 | Spectral conductivity in a parabolic band and the Kane band

Within the Boltzmann transport theory, the transport coefficient for a single parabolic band is expressed as

$$L_n = \int_0^\infty (E-\mu)^n \Sigma(E)\left(-\frac{\partial f_{FD}(E,T)}{\partial E}\right) dE, \tag{S1}$$

where the origin of the energy $E$ is set at the edge of the parabolic band. In isotropic systems, the spectral conductivity is given by

$$\Sigma(E) = \frac{e^2}{3} v^2(E)\tau(E)g(E), \tag{S2}$$

where $v(E), \tau(E)$, and $g(E)$ denote the group velocity, relaxation time, and density of states, respectively. For a parabolic system with the dispersion relation $E = \frac{\hbar^2 k^2}{2m^*}$, the group velocity and density of states are

$$v(E) = \frac{1}{\hbar}\frac{dE}{dk} = \frac{\hbar k}{m^*}, \tag{S3}$$

$$g(E) = \frac{1}{\pi^2} k^2 \frac{dk}{dE} = \frac{\sqrt{2} m^{*\frac{3}{2}}}{\pi^2 \hbar^3} E^{\frac{1}{2}}. \tag{S4}$$

By substituting these expressions into the definition of $\Sigma(E)$, the spectral conductivity for a parabolic band is obtained as

$$\Sigma(E) = \frac{2\sqrt{2}}{3}\frac{e^2 m^{*\frac{1}{2}}}{\pi^2 \hbar^3} E^{\frac{3}{2}} \tau(E). \tag{S5}$$

In the Kane model[1], by contrast, the band dispersion is described as

$$E(1+\alpha E) = \frac{\hbar^2 k^2}{2m^*}, \tag{S6}$$

where $\alpha = \frac{1}{E_g}$. The corresponding group velocity and density of states are given by

$$v(E) = \frac{\hbar}{m^*} \frac{k}{1 + 2\alpha E}, \tag{S7}$$

$$g(E) = \frac{1}{\pi^2} k^2 \frac{dk}{dE} = \frac{\sqrt{2} m^{*\frac{3}{2}}}{\pi^2 \hbar^3} (1 + 2\alpha E) \sqrt{E(1 + \alpha E)}. \tag{S8}$$

The resulting spectral conductivity in the Kane model is

$$\Sigma(E) = \frac{2\sqrt{2}}{3} \frac{e^2 m^{*\frac{1}{2}}}{\pi^2 \hbar^3} E^{\frac{3}{2}} \tau(E) \frac{(1 + \alpha E)^{\frac{3}{2}}}{1 + 2\alpha E}. \tag{S9}$$

Therefore, the ratio of spectral conductivity between the Kane model and the parabolic-band model is

$$\frac{\Sigma_{kane}}{\Sigma_{para}} = \frac{(1 + \alpha E)^{\frac{3}{2}}}{1 + 2\alpha E}. \tag{S10}$$

The energy dependence of this ratio is calculated for several $E_g$ values, as shown in Fig. S1. The ratio substantially deviates from unity when $E_g$ approaches zero; however, it remains approximately unity when $0.3 \text{ eV} \leq E_g \leq 0.5 \text{ eV}$, which corresponds to the band gap of PbTe at high temperatures[2]. This result indicates that approximating the $L$ band with a parabolic dispersion in PbTe does not significantly alter the calculated thermoelectric properties. This conclusion is consistent with the agreement between the calculated $zT$ values in Fig. 5b and the experimentally reported value of $zT \sim 2.5$ for Sr/Na-doped PbTe[3].

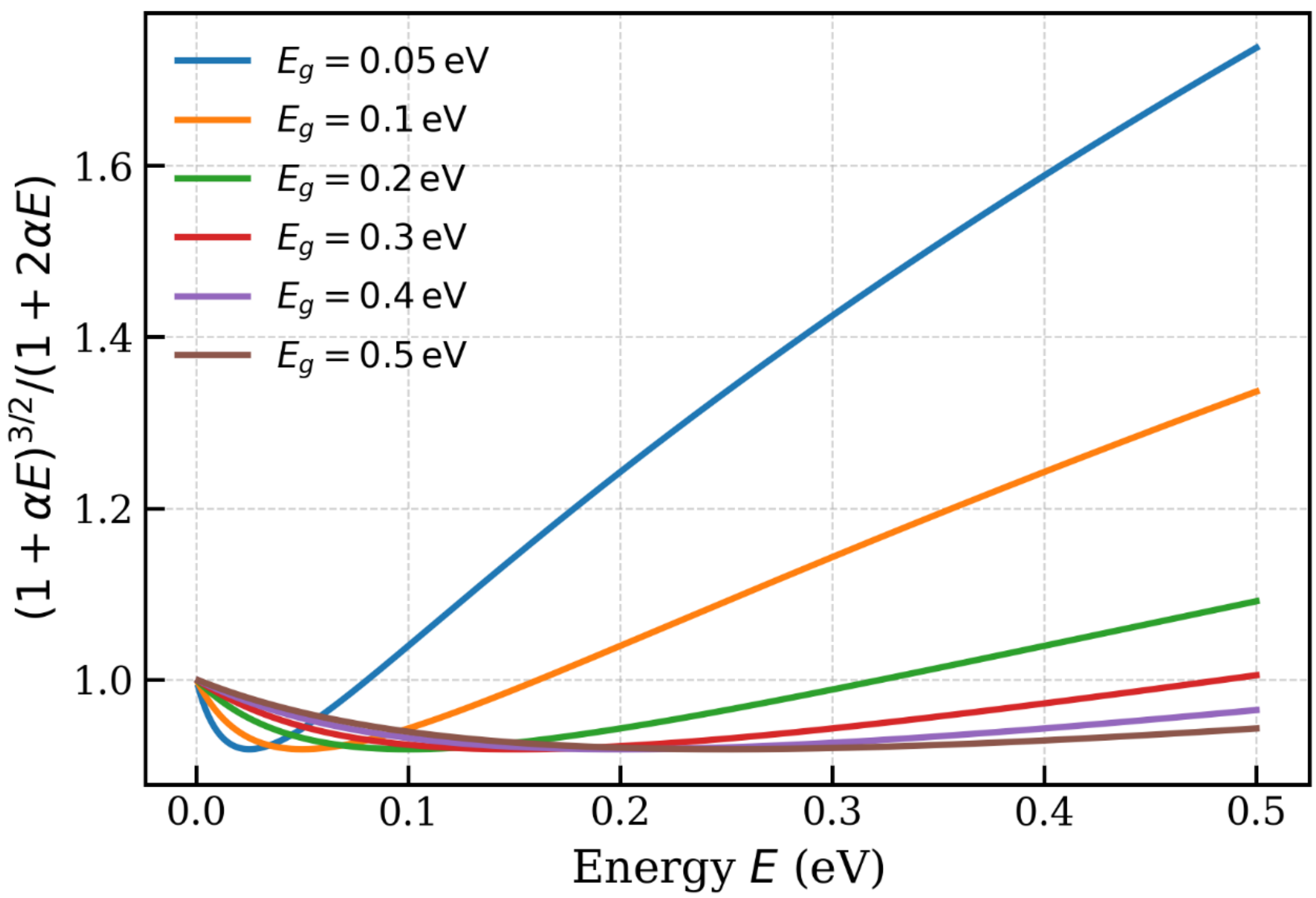

**Supplementary Fig. 1.** Energy dependence of the ratio $\Sigma_{kane}/\Sigma_{para}$ for several $E_g$ values.

**Supplementary Note 2. Optimal chemical potential when $\kappa_{lat} \gg \kappa_{el}$**

Using the expression in Eq. (S1), the electrical conductivity $\sigma$, Seebeck coefficient $S$, and electronic thermal conductivity $\kappa_{el}$ are given by

$$\sigma = L_0, \tag{S11}$$

$$S = -\frac{1}{eT}\frac{L_1}{L_0}, \tag{S12}$$

$$\kappa_{el} = \frac{1}{e^2 T}\left(L_2 - \frac{L_1^2}{L_0}\right) \tag{S13}$$

In the limit of $\kappa_{lat} \gg \kappa_{el}$, maximizing $zT = \sigma S^2 T/(\kappa_{lat} + \kappa_{el})$ reduces to maximizing the power factor $PF = \sigma S^2$.

$$PF = \frac{1}{e^2T^2}\frac{L_1^2}{L_0} \tag{S14}$$

The spectral conductivity $\Sigma(E)$ in isotropic systems is given by Eq. (S2). For a three-dimensional parabolic band under the constant relaxation time approximation, $v^2(E) \propto E$ and $g(E) \propto E^{1/2}$, and thus $\Sigma(E) = CE^{3/2}$, where $C$ is a constant value. When we introduce dimensionless variables

$$x = \frac{E}{k_BT}, \eta = \frac{\mu}{k_BT}, \tag{S15}$$

the transport coefficients can be written as

$$L_n = C(k_BT)^{\frac{3}{2}+n} I_n(\eta), \tag{S16}$$

with

$$I_n(\eta) = \int_0^{\infty} dx x^{\frac{3}{2}}\left(-\frac{\partial f}{\partial x}\right)(x-\eta)^n \tag{S17}$$

Using this expression, the power factor becomes

$$\mathrm{PF} = \frac{Ck_B^{\frac{7}{2}}T^{\frac{3}{2}}}{e^2}\frac{I_1^2(\eta)}{I_0(\eta)}. \tag{S18}$$

Therefore, the power factor depends only on the reduced chemical potential $\eta$. Using the spectral conductivity shown in Fig. S2a for an isotropic three-dimensional parabolic band within the constant relaxation time approximation, the optimal $\eta$ is found to be $\eta = -2.47$, as shown in Fig. S2b, leading to the optimal chemical potential $\mu = -2.47\ k_BT$.

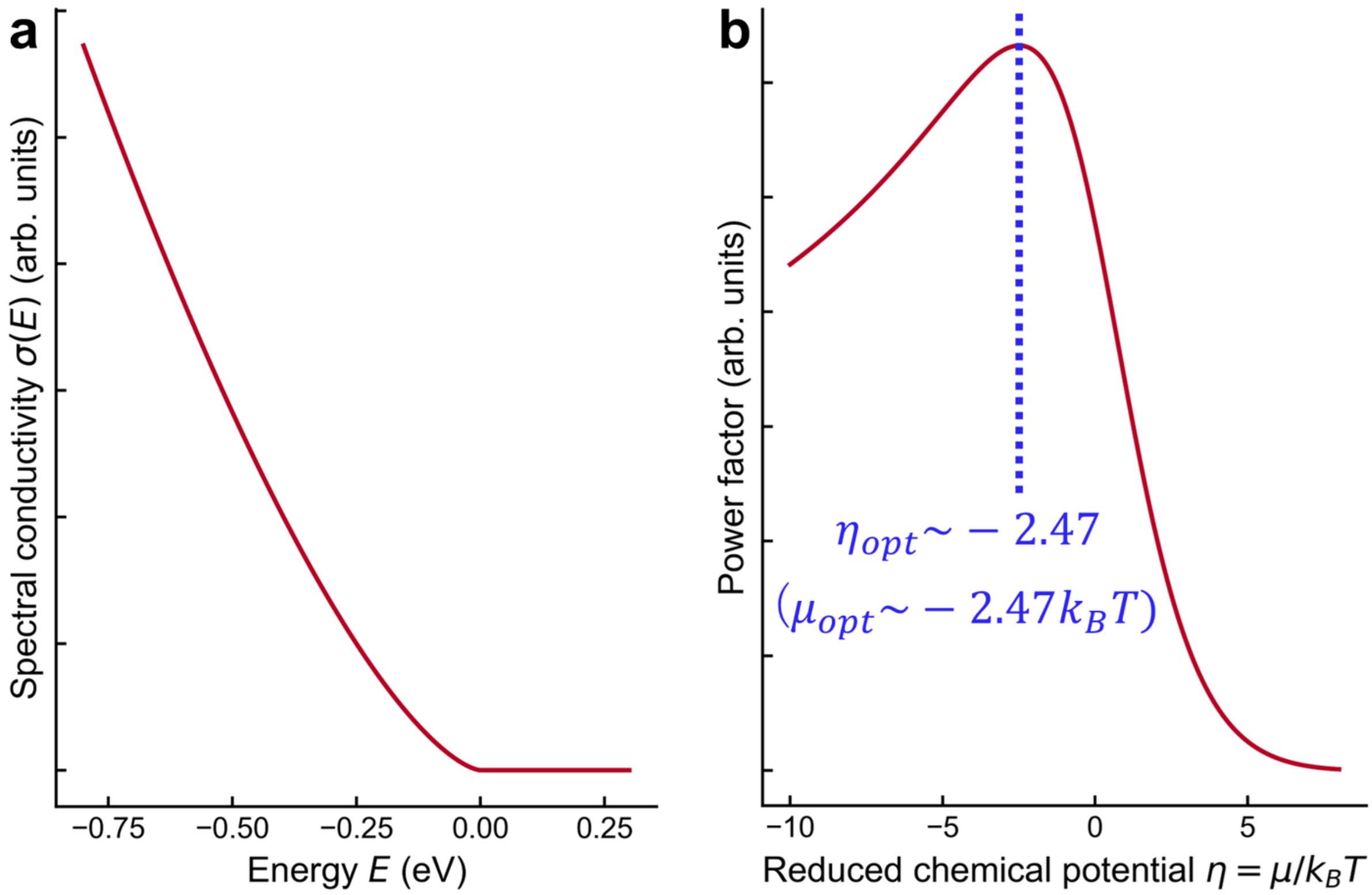

**Supplementary Fig. 2.** (a) Energy dependence of the spectral conductivity for an isotropic three-dimensional parabolic band within the constant relaxation time approximation. (b) Chemical potential dependence of the power factor within the constant relaxation time approximation.

**Supplementary Note 3 | Carrier distribution in Sr/Na-doped PbTe at $T = 900$ K**

Within the two-band model, the Hall coefficient of p-type PbTe can be expressed as

$$R_H = \frac{1}{e} \frac{n_L \mu_L^2 + n_\Sigma \mu_\Sigma^2}{(n_L \mu_\mathrm{L} + n_\Sigma \mu_\Sigma)^2}, \tag{S19}$$

where $n_i$ and $\mu_i$ represent the carrier concentration and mobility of band $i$, respectively. Since the band gap of PbTe exceeds $E_g = 0.5$ eV at high temperatures[2], and this value is larger than $5k_B T$ ($= 0.39$ $e$V at $T = 900$ K), the contribution of minority carriers can be neglected. Accordingly, the total carrier concentration $n_{tot}$ is set as a temperature-independent value.

The reduced Hall coefficient is then calculated as

$$\frac{R_H}{R_{H,0}} = \frac{1 - x + x\left(\frac{\mu_\Sigma}{\mu_L}\right)^2}{\left(1 - x + x\left(\frac{\mu_\Sigma}{\mu_L}\right)\right)^2}, \tag{S20}$$

where $R_{H,0}$ is the Hall coefficient at $T = 0$ K, and $x = n_\Sigma / n_{tot}$. Since all the carriers occupy the $L$ band at $T = 0$ K,

$$R_{H,0} = \frac{1}{e}\frac{n_L \mu_L^2}{(n_L \mu_L)^2} = \frac{1}{e n_{tot}}. \tag{S21}$$

Figure S3 shows the $x$ dependence of $\frac{R_H}{R_{H,0}}$, together with the ratio of the electrical conductivity from the $\Sigma$ band to the total electrical conductivity, $\sigma_\Sigma / \sigma_{tot}$. The mobility ratio $\mu_\Sigma / \mu_\mathrm{L} = 0.075$, determined in a previous study of Sr/Na-codoped PbTe[4], was used in the calculation. In the previous experimental study of Sr/Na-codoped PbTe, $R_H / R_{H,peak} \sim 0.2$ was reported at $T = 825$ K[3]. It corresponds to $x = n_\Sigma / n_{tot} \sim 1.0$ in Fig. S3, indicating that the total carrier concentration as well as total electrical conductivity is dominated by the $\Sigma$ band at $T = 900$ K.

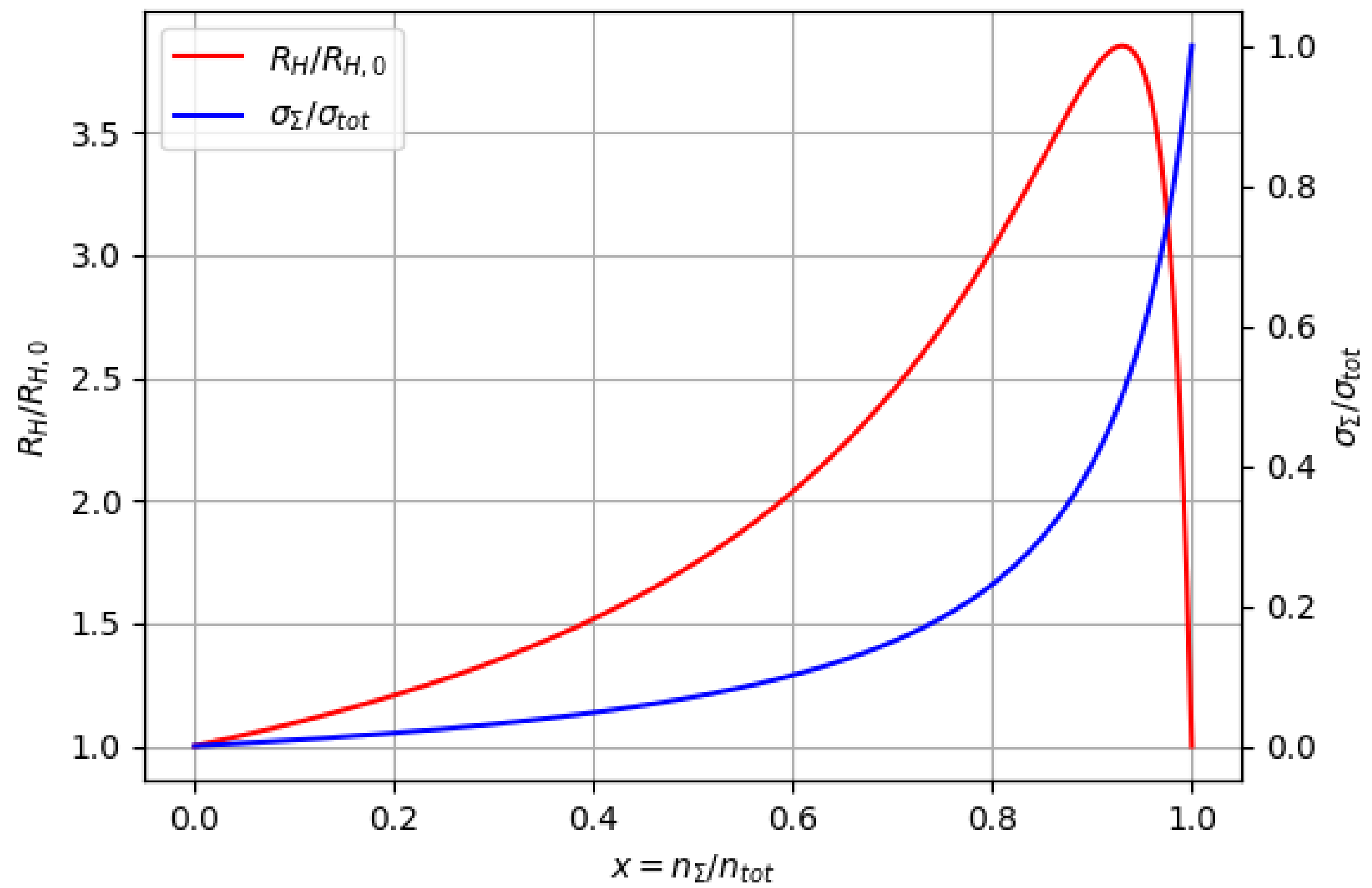

**Supplementary Fig. 3.** $x(= n_\Sigma/n_{tot})$ dependence of $R_H/R_{H,0}$ and $\sigma_\Sigma/\sigma_{tot}$ for $\mu_\Sigma/\mu_\mathrm{L} = 0.075$.

**Supplementary Note 4. Chemical potential dependence of $zT$ with an energy-dependent relaxation time**

Here, we calculate thermoelectric properties assuming an energy-dependent relaxation time of the form $\tau(E) = \tau_0 E^\gamma$, with $\gamma = -0.5$ and $\gamma = 0.5$. Since the spectral conductivity for a three-dimensional parabolic band with constant relaxation time ($\gamma = 0$) is given by $\Sigma_i \propto \left(E_{i,v} - E\right)^{\frac{3}{2}}$, the energy dependence of the spectral conductivity is modified to $\Sigma_i \propto E_{i,v} - E$ for $\gamma = -0.5$, and $\Sigma_i \propto \left(E_{i,v} - E\right)^2$ for $\gamma = 0.5$. The band model for $\gamma = -0.5$, energy dependence of the spectral conductivity $\Sigma$, the chemical potential dependence of the electrical conductivity $\sigma$, Seebeck coefficient $S$, and electronic thermal conductivity $\kappa_{el}$ are presented in Figs. S4a, S4b, S4c, S4d, S4e, and S4f, respectively. The same calculations were performed for $\gamma = 0.5$, and the results are shown in Figs. S4g-l. Although the absolute values of $zT$ depend on the exponent $\gamma$, the highest $zT$ value is consistently obtained when $\Delta E = 0$ eV. Similar calculations were performed to investigate the dependence of $zT$ on the heavy-band DOS effective mass $m^*_{DOS,H}$ for $\gamma = -0.5$ and $\gamma = 0.5$, as shown in Figs. S4m-r and S4s-x, respectively. In each case, $zT$ is enhanced when $r_m = m^*_{DOS,H}/m^*_{DOS,L}$ is large.

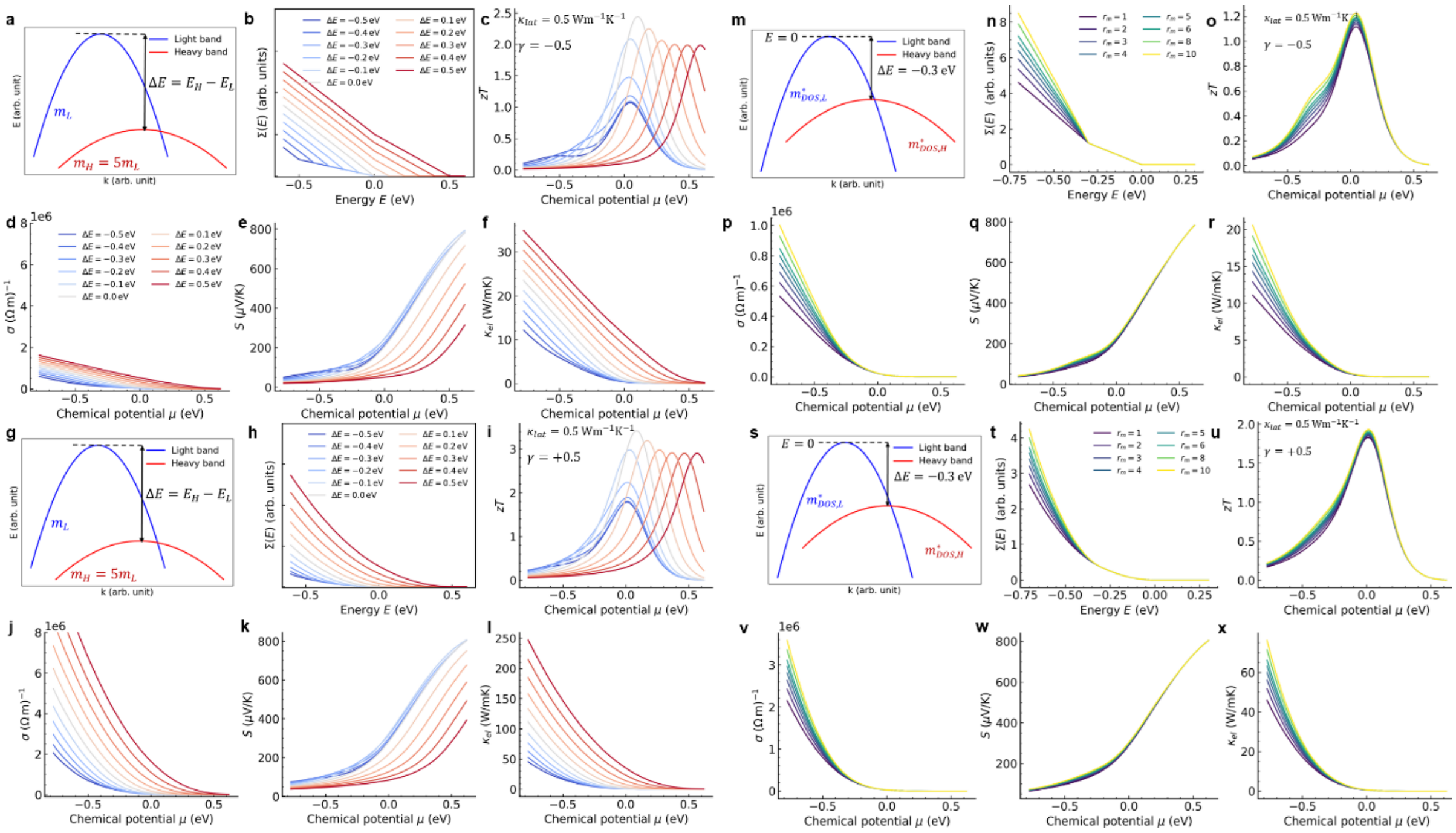

**Supplementary Fig. 4.** (a) Schematic illustrations of the band structure, (b) spectral conductivity, and the chemical potential dependence of (c-f) thermoelectric properties for $\gamma = -0.5$ for several $\Delta E$ values. (g-l) Similar calculations for $\gamma = 0.5$. (m) Schematic illustrations of the band structure, (n) spectral conductivity, and the chemical potential dependence of (o-r) thermoelectric properties for $\gamma = -0.5$ for several $m^*_{DOS,H}$ values. (s-x) Similar calculations for $\gamma = 0.5$.

**Supplementary Note 5. Determination of the chemical potential $\mu$ from the carrier**

**concentration $\boldsymbol{n}$**

As discussed in the main text, the relation between the carrier concentration $n$ and chemical potential $\mu$ in PbTe at high temperatures is expressed as

$$n = \frac{\left(2m_{d,\Sigma}^{*}\right)^{3/2}}{2\pi^2\hbar^3} N_\Sigma \int_{-\infty}^{E_{\Sigma,\mathrm{v}}} \frac{\left(E_{\Sigma,\mathrm{v}} - E\right)^{1/2}}{1 + \exp\frac{\mu_{ref} - E}{k_B T}} dE.$$

Since $N_\Sigma = 12$ and $m_{d,\Sigma}^{*} = 0.34\, m_0$ for the $\Sigma$ band in PbTe[5], the relation between the chemical potential $\mu$ and the hole concentration $p$ is expressed as shown in Fig. S5. The total hole concentration of Sr/Na-doped PbTe at the solubility limit is estimated to be $n \approx 1.3 \times 10^{20}\ \mathrm{cm}^{-3}$[4]. Thus, it corresponds to $\mu \sim +60$ meV, implying that $\mu$ lies above the band edge. Accordingly, the reference chemical potential in our calculations is $\mu_{ref} = +60$ meV.

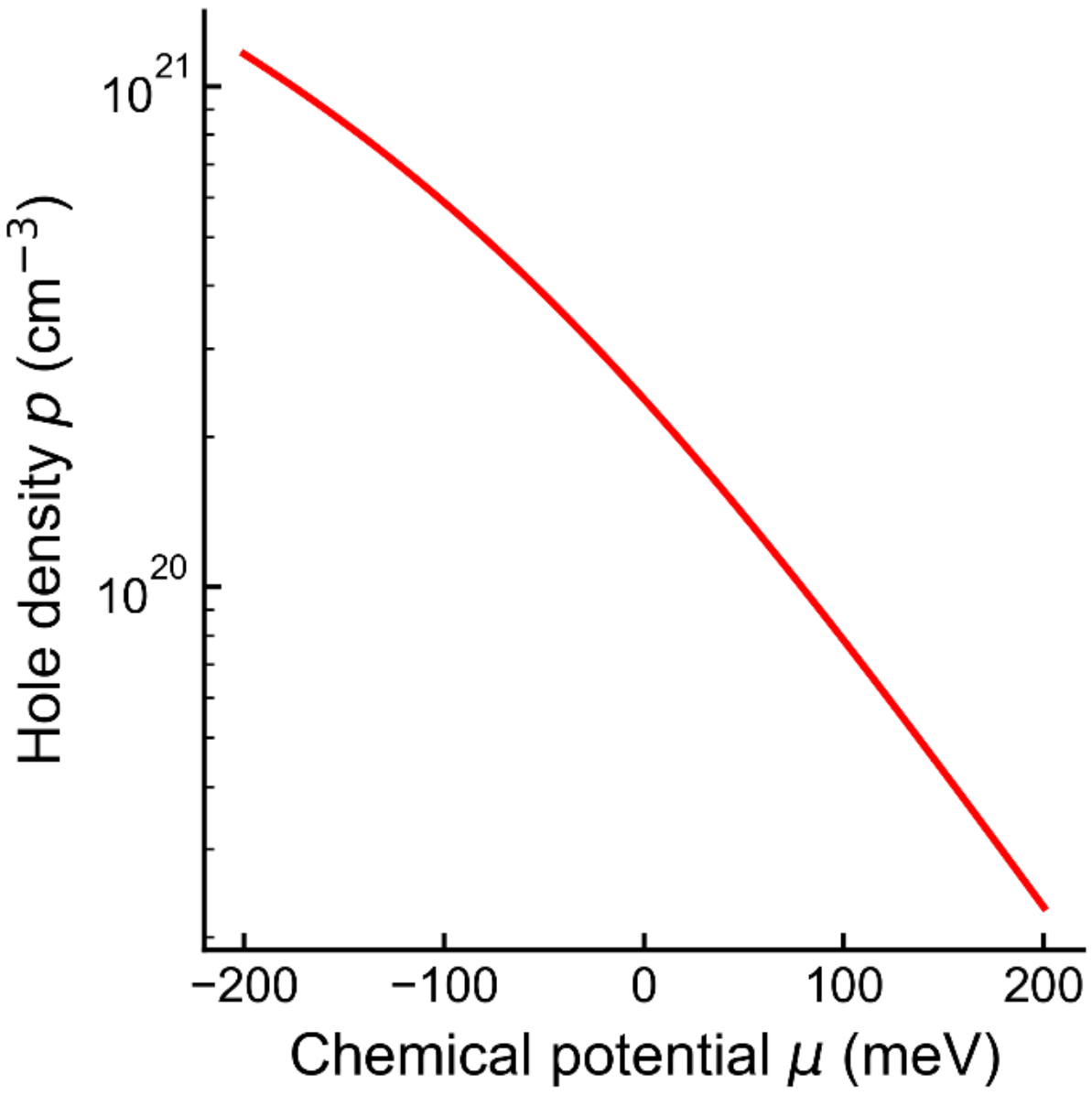

**Supplementary Fig. 5.** Relation between the chemical potential $\mu$ and the hole concentration $n$ at $T = 900$ K for $N_\Sigma = 12$ and $m_{d,\Sigma}^{*} = 0.34\, m_0$.

**Supplementary Note 6. $\boldsymbol{\Delta E}$ dependences of $\boldsymbol{\kappa_{el}}$ in the band-converged system**

Figure S6 shows the $\Delta E$ dependences of $\kappa_{el}$ in the band-converged system. Detailed descriptions of the calculation methods are provided in the Methods section of the main text (4.4 Normalization factor of spectral conductivity $\alpha$ and band parameters).

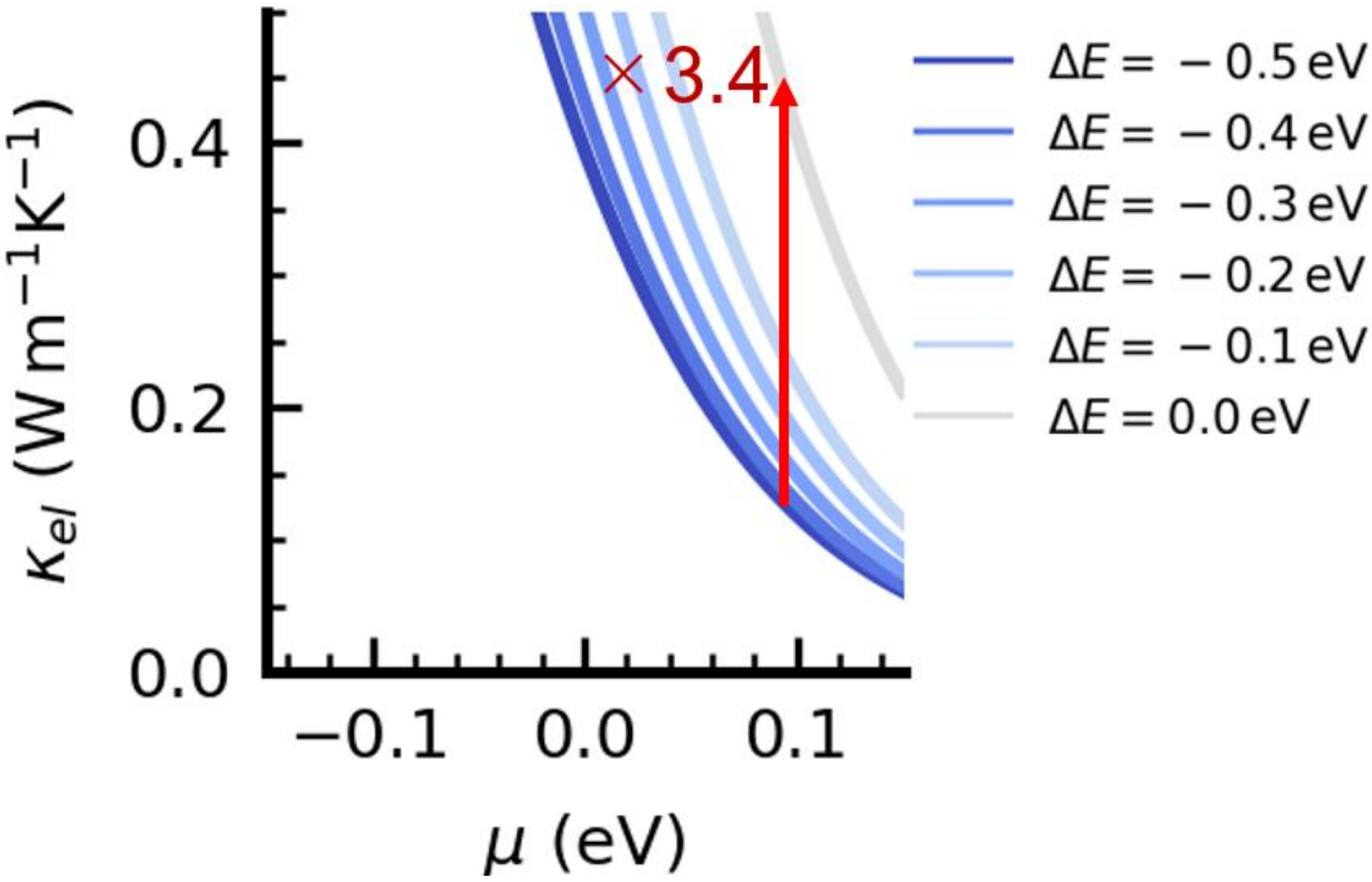

**Supplementary Fig. 6.** Chemical potential dependences of $\kappa_{el}$ at several $\Delta E$ values.

**Supplementary Note 7. Minimum $E_g$ required to suppress the bipolar effect**

In Fig. 4e in the main text, we show that the minimum $E_g$ required to suppress the bipolar effect in $S$ is $5k_BT$ while that in $\kappa_{el}$ is $7k_BT$. We defined the minimum band gap for $S$ as the gap at which the bipolar-induced reduction in $|S|$ is limited to 5% or less. Similarly, for $\kappa_{el}$, we defined the minimum band gap as the gap at which the bipolar-induced enhancement of $\kappa_{el}$ is limited to 10% or less.

In the main text, we investigated the $E_g$ dependence of $\sigma, S,$ and $\kappa_{el}$ for a parabolic band within the constant relaxation time approximation. In real materials, the relaxation time $\tau$ can exhibit an energy dependence. Therefore, when the spectral conductivity is expressed as $\Sigma(E) = \Sigma_0 E^{\beta}$, the exponent $\beta$ may take values other than 1.5. Figure S7 shows the $E_g$ dependence of $\sigma, S$ and $\kappa_{el}$ for different $\beta$ values, which is normalized to the corresponding values calculated for $E_g = 10k_BT$ and $\mu = 0$ (band edge). The bipolar effect is more significant at $\beta = 1.0$ than at $\beta = 2.0$ because the spectral conductivity increases more sharply near the band edge of the minority carriers. However, the $\beta$ dependence is relatively weak. This can be understood from the fact that the window function in Fig. 3 decays exponentially in the energy range where $|E - \mu|/k_BT$ is sufficiently large.

Lastly, we note that the $7k_BT$ criterion for $\kappa_{el}$ may be too stringent for practical use, because real materials also have a lattice thermal conductivity $\kappa_{lat}$, which is typically comparable to $\kappa_{el}$ in high-performance thermoelectric materials. For example, even if $\kappa_{el}$ is increased by 10% due to the bipolar effect, the corresponding increase in the total thermal conductivity $\kappa$ would be only about 5%.

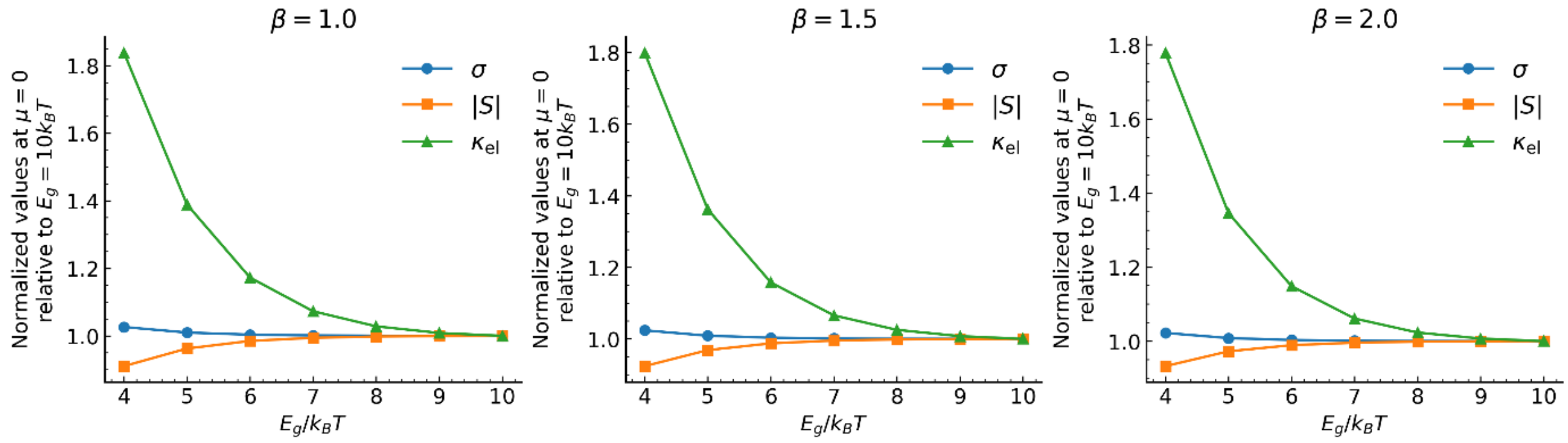

**Supplementary Fig. 7.** $E_g$ dependences of $\sigma, S$, and $\kappa_{el}$ for $\beta = 1.0$, 1.5, and 2.0.

## Supplementary References